# On Flow-Induced Diffusive Mobile Molecular Communication: First Hitting Time and Performance Analysis

Neeraj Varshney*, *Student Member, IEEE,* Werner Haselmayr, *Member, IEEE,* and Weisi Guo, *Member, IEEE*

*Abstract*—This work considers the problem of flow-induced diffusive molecular communication under various mobility conditions such as (i) both transmitter (TX) and receiver (RX) nanomachines are mobile, (ii) TX is mobile and RX is fixed, and (iii) TX is fixed and RX is mobile. Closed-form expressions for the probability density function (PDF) of the first hitting time under the aforementioned mobile scenarios are derived, by characterizing the movement of the nanomachines and information molecules using Brownian motion with positive drift. The derived PDF expressions are validated through particle-based simulations. Based on these results, the performance of molecular communication with on-off keying (OOK) modulation in flow-induced diffusive channels is investigated. In particular, closed-form expressions for the probabilities of detection and false alarm with optimal Likelihood ratio test (LRT) based decision rule, probability of error, and the capacity in the presence of inter-symbol interference, counting errors, and noise from the other sources are derived. Simulation results are presented to verify the theoretical results and to yield insights into the system performance for different mobility conditions.

## I. INTRODUCTION

MOLECULAR communication broadly defines the information exchange using chemical signals [1]. This new paradigm is a promising candidate for the communication between nanomachines due to its ultra-high energy efficiency and bio-compatibility [2]. The envisaged applications of molecular communication are in the area of biomedical, environmental and industrial engineering [3], [4]. Currently, molecular communication research can be split into three areas: 1) *Living system modelling* aims at gaining more insights into molecular communication processes occurring in biological live system using techniques originating from communications engineering. For example, in [5] information theory was used to quantify the information exchange in protein structures. 2) *Living systems interface* aims to control the behavior of biological systems. For example, in [6] the redox modality is used to connect synthetic biology to electronics. 3) *Artifical molecular communication* focuses on the design, fabrication and testing of human-made molecular communication systems. For example, an inexpensive experimental platform was presented in [7], which can be used to simulate different environments like the cardiovascular system. Moreover, a chemical communication between gated nanoparticles was recently shown in [8].

### A. Motivation and Related Work

One promising future application for artificial molecular communication systems is chrono drug-delivery [9], whereby nanomachines in one part of the body senses an event and communicates to drug delivery nanomachine in another part of the body using molecular signaling via the circulation system. This is akin to hormone signaling between spatially far apart organs (the endocrine system). Fig. 1 shows a concept system, whereby two nanomachines are separated by a significant distance in a blood circulatory system with advection flow. Both the nanomachines themselves and the molecular messages undergo a combined diffusion-advection transport process, whereby neither the diffusion nor the advection process is overly dominant. As such, analyzing the first hitting process, optimal decision thresholds, communication reliability and capacity performance is of interest, especially with active receivers.

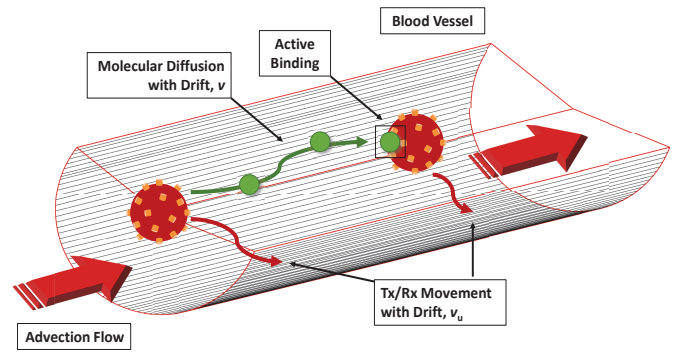

Fig. 1. Mobile molecular communications in diffusion-advection blood flow channels.

Mobile molecular communication has been considered in [10]–[20]. A clock synchronization scheme between a fixed and a mobile nanomachine is proposed in [10]. A protocol for mitigating inter-symbol interference (ISI) for diffusion based mobile molecular communication is presented in [11]. Similarly, the work in [12] investigates different coding strategies for mitigating transposition errors for flow-induced diffusion mobile molecular communication. In the works [10]–[12] the mobility of the nanomachines is modeled through a time-varying distance (e.g., distance increases with equal difference [10]). Mobile bacteria networks are proposed in [13], applying an independently and identically distributed mobility model. That is, in each time slot a network node chooses its new position independently and identically distributed over the entire network. Mobile ad-hoc nano-networks are considered in [14] and [15], where the mobile nanomachines freely diffuse in a three-dimensional (3D) environment via Brownian motion. In [14] the information transfer is accomplished through neurospike communication after collision and adhesion of the nodes and in [15] it based on Förster resonance energy transfer if the nodes are in close proximity of each other. A mathematical model of non-diffusive mobile molecular communication networks is presented in [16], describing the two-dimensional (2D) mobility of the nanomachines through the Langevin equations. In [17] the channel impulse response (CIR) for diffusive mobile molecular communication in a 3D environment is presented, assuming that only the receiver (RX) is mobile and a fully absorbing receiver. The CIR is obtained by modifying the diffusion coefficient of the information molecules. In [18] the first hitting time distribution for a one-dimensional (1D) diffusion channel without drift is derived, taking the mobility of the transmitter (TX) and RX into account. Further, using the first hitting time distribution [18], the performance of diffusive mobile molecular communication system without drift has been recently analyzed in [19]. In [20] the CIR for a mobile molecular communication system is presented, considering the mobility of both TX and RX as well as the impact of flow. However, in contrast to [17]–[19] TX and RX are assumed to be as passive. It is important to note that the information transfer in [17]–[20] is accomplished through freely diffusing particles, whereas in [14]–[16] other mechanisms (e.g., Förster resonance energy transfer) are exploited.

*Corresponding author.
N. Varshney is with the Department of Electrical Engineering, Indian Institute of Technology Kanpur, India (email: neerajv@iitk.ac.in).
W. Haselmayr is with the Johannes Kepler University Linz, Austria (email: werner.haselmayr@jku.at).
W. Guo is with the School of Engineering, University of Warwick, United Kingdom (email: weisi.guo@warwick.ac.uk).

### B. Contributions

In this work, we derive the first hitting time distributions for mobile molecular communication in flow-induced diffusion channels and provide a comprehensive performance analysis for such a system. In particular this work makes the following contributions:

1) Closed-form expressions for the first hitting time distribution in flow-induced diffusion channels are derived, considering various mobility conditions such as (i) both TX and RX are mobile, (ii) TX is mobile and RX is fixed, and (iii) TX is fixed and RX is mobile. The RX is modeled as fully absorbing RX [21], which is in contrast to [20] where only a passive RX is investigated. The derived expressions are validated through particle-based simulations. To the best of our knowledge, this is the first time that the analysis for the first hitting time distribution for flow-induced diffusion channels with mobile TX and mobile active RX is reported.

2) This work also provides a comprehensive performance analysis for molecular communication between mobile nanomachines in flow-induced diffusion channels, considering multiple-source interference (MSI), ISI and counting errors at the RX. In particular, an optimal decision rule and an adaptive decision threshold are derived based on the optimal Likelihood ratio test (LRT). Unlike the decision threshold presented in [20, (42)], the decision threshold in this work does not require knowledge on the previously transmitted bits, which is not known in practice. Based on the derived optimal decision rule the detection performance in terms of receiver operating characteristic, error probability, and the channel capacity are investigated. Both analytical expression and computer simulations for the aforementioned metrics are presented to develop several important and interesting insights into the system performance under various mobile scenarios.

### C. Organization

The rest of the paper is organized as follows. Section II presents the system model for diffusive molecular communication with mobile TX and RX under drift channel. The derivations of first hitting time distribution under various mobile scenarios, along with validation using the particle-based simulations, are presented in Section III. This is followed by the comprehensive performance analysis in Section IV where the analytical results for the optimal decision rule at the RX, probabilities of detection and false alarm, the total error probability and the channel capacity are derived. Section V demonstrates the performance analysis through computer simulations under various mobile conditions. Finally, Section VI provides concluding remarks.

## II. SYSTEM MODEL FOR DIFFUSIVE MOBILE MOLECULAR COMMUNICATION

Consider a diffusive molecular communication system with TX and RX in motion while communicating in a semi-infinite 1D fluid medium with drift. Moreover, this work assumes constant temperature and viscosity. TX and RX are placed on a straight line at a certain distance from each other and the movement of both nanomachines is modeled as a 1D Gaussian random walk [18]. Similar to [22]–[24] and the references therein, both nanomachines are assumed to be synchronized in time. The channel is divided into time slots of duration $T$ as shown in Fig. 2, where the $j$th slot is defined as the time period $[(j-1)T, jT]$ with $j \in \{1, 2, \cdots\}$. At the beginning of each time slot, the TX either emits the same type of molecules in the propagation medium with prior probability $\beta$ for transmission of information symbol 1 or remains silent for transmission of information symbol 0. Let $\mathcal{Q}[j]$ denote the number of molecules released by the TX for information symbol $x[j] = 1$ at the beginning of the $j$th slot. The molecular propagation from the TX to RX occurs via Brownian Motion with diffusion coefficient denoted by $D_\text{m}$. Similar to [23], [24] and the references therein, it is also assumed that the transmitted molecules do not interfere or collide with each other. Once the molecules reach the RX, they are immediately absorbed, followed by detection of the transmitted information based on the number of molecules received.

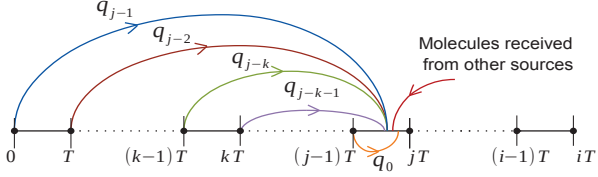

Fig. 2. Diffusion based molecular communication over time slotted channel, where $[(j-1)T, jT]$ is the current slot and $q_0$ is the probability that a molecule reaches RX within the current slot.

Due to the stochastic nature of the diffusive channel, the times of arrival at the RX, of the molecules emitted by the TX, are random in nature and can span multiple time slots. Let $q_{j-k}$ denote the probability that a molecule transmitted in slot $k \in \{1, 2, \cdots, j\}$ arrives in time slot $j$ and can be obtained as

$$q_{j-k} = \int_{(j-k)T}^{(j-k+1)T} f_{T_h}(t; k) dt, \quad (1)$$

where $f_{T_h}(t; k)$ denotes the probability density function (PDF) of first hitting time at the RX. The following section derives the closed-form expression for the PDF of first hitting time considering the various mobility conditions of TX and RX in flow-induced diffusive channel with drift.

## III. First Hitting Time

This section begins by considering the first hitting time distribution for fixed TX and RX under flow-induced diffusive channel. Then, the variation of the Euclidean distance when TX and RX are moving is characterized, which is subsequently used to derive the first hitting time distribution for various mobility scenarios. Finally, the derived expressions are validated through particle-based simulations.

The first hitting time for flow-induced diffusion channels with fixed TX and RX follows an inverse Gaussian (IG) distribution with its PDF given by [25]

$$f_{T_h}(t) = \frac{r_0}{\sqrt{4\pi D_m t^3}} \exp\left(-\frac{(r_0 - v^\star t)^2}{4 D_m t}\right), \quad t > 0, \quad (2)$$

where the Euclidean distance between TX and RX is given by $r_0 = \sqrt{(x_{0,\mathrm{rx}} - x_{0,\mathrm{tx}})^2}$, and $x_{0,\mathrm{tx}}$ and $x_{0,\mathrm{rx}}$ denote the $x$-coordinate position of TX and RX, respectively. The velocity $v^\star$ is given by

$$v^\star = \mathrm{sgn}(x_{0,\mathrm{rx}} - x_{0,\mathrm{tx}})v = \mathrm{sgn}(d_0)v, \quad (3)$$

where $\mathrm{sgn}(\cdot)$ denotes the sign function, $d_0 = x_{0,\mathrm{rx}} - x_{0,\mathrm{tx}}$ and $v$ corresponds to the positive flow velocity from TX to RX. The definition in (3) allows to model molecular communication with both positive and negative flow[1].

In case of no flow, i.e., $v = 0$, the IG distribution in (2) turns into a Lévy distribution given by

$$f_{T_h}(t) = \frac{r_0}{\sqrt{4\pi D_m t^3}} \exp\left(-\frac{r_0^2}{4 D_m t}\right), \quad t > 0. \quad (4)$$

[1]Please note that the first hitting time in case of a negative flow can be modeled through an IG distribution with negative flow velocity $v$ [25].

### A. Distance Variation between TX and RX

In case of mobile TX and/or RX the diffusion coefficients of the mobile TX and RX are denoted by $D_{\mathrm{tx}}$ and $D_{\mathrm{rx}}$, respectively. Each of them is either fixed or moves with the flow $v$, i.e., the velocities of TX and RX are given by $v_{\mathrm{tx}}, v_{\mathrm{rx}} \in \{0, v\}$. This work assumes that the movement of TX and RX does not disrupt the propagation of the information molecules and that TX and RX can pass each other. The Euclidean distance between TX and RX at the beginning of the $k$th time slot, i.e., at time $\tau = kT$, is given by

$$r_k = \sqrt{(x_{k,\mathrm{rx}} - x_{k,\mathrm{tx}})^2} = \sqrt{d_k^2}, \quad (5)$$

with $d_k = x_{k,\mathrm{rx}} - x_{k,\mathrm{tx}}$ and $x_{k,u}$, $u \in \{\mathrm{tx}, \mathrm{rx}\}$, denotes the $x$-coordinate position of TX and RX at time $\tau = kT$. It is important to note that the Euclidean distance used in (2) corresponds to the Euclidean distance in (5) at time $\tau = 0$. The positions of TX and RX at time $\tau = kT$ considering their mobility in drift channels can be defined as follows [1]

$$\begin{aligned} x_{k,u} &= x_{k-1,u} + \Delta x_u \\ &= x_{0,u} + \sum_{i=1}^{k} \Delta x_u, \quad u \in \{\mathrm{tx}, \mathrm{rx}\}, \end{aligned} \quad (6)$$

where $x_{0,u}$, $u \in \{\mathrm{tx}, \mathrm{rx}\}$, denote the initial position of TX and RX and $\Delta x_u = v_u T + \tilde{\Delta} x_u$. The random displacement $\tilde{\Delta} x_u$ follows a Gaussian distribution with zero mean and variance $\sigma_u^2 = 2 D_u T$, i.e., $\tilde{\Delta} x_u \sim \mathcal{N}(0, 2D_u T)$. Thus, the position and the distance of TX and RX at time $\tau = kT$ follows a Gaussian distribution

$$x_{k,u} \sim \mathcal{N}(\bar{x}_{k,u}, k\sigma_u^2), \quad (7)$$
$$d_k \sim \mathcal{N}(\bar{d}_k, \sigma_k^2), \quad (8)$$

with $\bar{x}_{k,u} = x_{0,u} + v_u kT$, $\bar{d}_k = d_0 + kT(v_{\mathrm{rx}} - v_{\mathrm{tx}})$, and $\sigma_k^2 = k(\sigma_{\mathrm{tx}}^2 + \sigma_{\mathrm{rx}}^2) = 2kT D_{\mathrm{tot}}$, with $D_{\mathrm{tot}} = D_{\mathrm{tx}} + D_{\mathrm{rx}}$. The distribution of the Euclidean distance $r_k$ defined in (5) is provided in the following theorem.

**Theorem 1.** *Since the distance $d_k$ follows a Gaussian distribution, the Euclidean distance $r_k = \sqrt{d_k^2}$ follows a scaled noncentral chi distribution with one degree of freedom. The PDF is given by*

$$f_{r_k}(r_k) = \frac{1}{\sqrt{\sigma_k^2}} \left[ \exp\left(-\frac{\left(\frac{r_k}{\sqrt{\sigma_k^2}}\right)^2 + \lambda^2}{2}\right) \sqrt{\lambda \frac{r_k}{\sqrt{\sigma_k^2}}} I_{-\frac{1}{2}}\left(\lambda \frac{r_k}{\sqrt{\sigma_k^2}}\right) \right], \quad (9)$$

*with noncentrality parameter $\lambda = \sqrt{\bar{d}_k^2 / \sigma_k^2}$ and the modified Bessel function of the first kind $I_{-\frac{1}{2}}(y)$ [26].*

*Proof.* The Euclidean distance $r_k$ (cf. (5)) can be written as

$$r_k = \sqrt{\frac{\sigma_k^2}{\sigma_k^2}} \sqrt{d_k^2} = \sqrt{\sigma_k^2} \tilde{r}_k, \quad (10)$$

where $\tilde{r}_k = \sqrt{d_k^2 / \sigma_k^2}$ denotes the scaled Euclidean distance. Since $d_k$ follows a Gaussian distribution (cf. (8)), the scaled

Euclidean distance $\tilde{r}_k$ follows a noncentral chi distribution with one degree of freedom [27]. Its PDF is given by

$$f_{\tilde{r}_k}(\tilde{r}_k) = \exp\left(-\frac{\tilde{r}_k^2 + \lambda^2}{2}\right) \sqrt{\lambda \tilde{r}_k}\, I_{-\frac{1}{2}}(\lambda \tilde{r}_k), \quad (11)$$

with the noncentrality parameter $\lambda = \sqrt{\bar{d}_k^2/\sigma_k^2}$. The Euclidean distance $r_k$ is defined by $\sqrt{\sigma_k^2}\tilde{r}_k$ and, thus, the corresponding PDF can be expressed as

$$f_{r_k}(r_k) = \frac{1}{\sqrt{\sigma_k^2}} f_{\tilde{r}_k}\left(\frac{r_k}{\sqrt{\sigma_k^2}}\right), \quad (12)$$

where the PDF $f_{\tilde{r}_k}(\cdot)$ is defined in (11). $\square$

### B. First Hitting Time Distribution

Similar to [18], [20] this work distinguishes between three cases, based on the mobility of TX and RX, for deriving the first hitting time distribution of molecules released at time $\tau = kT$.

*1) Fixed TX and mobile RX:* In this case, TX is fixed and only RX is moving along with the information molecules with flow $v$, i.e., $v_{\text{tx}} = D_{\text{tx}} = 0$, $D_{\text{rx}} \neq 0$, and $v_{\text{rx}} = v$. The first hitting time distribution can be obtained through considering the relative motion between the information molecules and the RX. This can be accomplished by an effective diffusion coefficient $D_{\text{eff}} = D_{\text{m}} + D_{\text{rx}}$ [28] and an effective flow velocity $v_{\text{eff}} = v - v_{\text{rx}}$. Since the RX is moving with flow $v$ the effective velocity $v_{\text{eff}}$ is zero. Thus, the first hitting time for fixed TX and mobile RX follows a Lévy distribution as defined in (4), with $r_0$ and $D_{\text{m}}$ substituted by $\bar{d}_k = d_0 + kTv$ and $D_{\text{eff}} = D_{\text{m}} + D_{\text{rx}}$, respectively.

*2) Mobile TX and fixed RX:* In this case, TX is moving along with the information molecules with flow $v$ and RX is fixed, i.e., $v_{\text{rx}} = D_{\text{rx}} = 0$, $D_{\text{tx}} \neq 0$, and $v_{\text{tx}} = v$. The PDF of the first hitting time can be computed as follows

$$f_{T_h}(t;k) = \int_{r_k=0}^{\infty} f_{T_h|r_k}(t|r_k) f_{r_k}(r_k) \mathrm{d}r_k, \quad (13)$$

where $f_{r_k}(r_k)$ is defined in (9) and $f_{T_h|r_k}(t|r_k)$ is given by

$$f_{T_h|r_k}(t|r_k) = \frac{r_k}{\sqrt{4\pi D_{\text{m}} t^3}} \exp\left(-\frac{(r_k - v^\star t)^2}{4 D_{\text{m}} t}\right), \quad t > 0, \quad (14)$$

with $v^\star = \text{sgn}(\bar{d}_k)v$. The variable $k$ denotes the time slot in which the molecules are released; in the $k$th time slot the molecules are released at time $\tau = kT$. The variable $t$ denotes the relative hitting time at the RX for a fixed time slot number $k$. The closed-form solution for (13) is given in (15) and the corresponding proof can be found in Appendix A. In (15), the PDF $f_{T_h}(t)$ corresponds to the first hitting time PDF for fixed TX and RX in (2) and the error function $\text{erf}(x)$ is defined by $\text{erf}(x) = \frac{2}{\sqrt{\pi}} \int_0^x \exp(-t^2) \mathrm{d}t$ [26, Eq. (8.250.1)].

*3) Mobile TX and RX:* In this case, TX and RX are moving along with the information molecules with flow $v$, i.e., $v_{\text{tx}} = v_{\text{rx}} = v$, $D_{\text{tx}} \neq 0$, and $D_{\text{rx}} \neq 0$. The PDF of the first hitting time is given by (15), with $D_{\text{m}}$, $D_{\text{tx}}$ and $v^\star = \text{sgn}(\bar{d}_k)v$ substituted by $D_{\text{eff}}$, $D_{\text{tot}}$ and $v^\star = \text{sgn}(\bar{d}_k)v_{\text{eff}} = \text{sgn}(\bar{d}_k)(v - v_{\text{rx}}) = 0$, respectively. It can be easily verified that (15) for mobile TX and RX in drift channel reduces to

$$f_{T_h}(t;k) = \frac{\sqrt{kT D_{\text{tot}} D_{\text{eff}}}}{\pi(kT D_{\text{tot}} + D_{\text{eff}} t)\sqrt{t}} \exp\left(\frac{-d_0^2}{4kT D_{\text{tot}}}\right)$$
$$+ f_{T_h}\left(t + \frac{kT D_{\text{tot}}}{D_{\text{eff}}}\right)$$
$$\times \text{erf}\left(\frac{d_0}{2}\sqrt{\frac{D_{\text{eff}} t}{kT D_{\text{tot}}(D_{\text{eff}} t + kT D_{\text{tot}})}}\right), \quad (16)$$

which is equivalent to the first hitting time PDF [18, Eq. (6)] for diffusion channels without flow and mobile TX and RX. This is because TX, RX and the information molecules are all moving with the same flow $v$.

Further, it can also be easily verified that in case of fixed TX and mobile RX, (15) reduces to a Lévy distribution as defined in (4) with the parameters $\bar{d}_k$ and $D_{\text{eff}}$.

### C. Evaluation of the First Hitting Time Distribution

For the evaluation of the first hitting time PDF, this work adopts the parameters used in [18], [20]: Without loss of generality this work assumes that TX and RX are located at the $x$-coordinates $x_{0,\text{tx}} = 0$ and $x_{0,\text{rx}} = 10^{-6}$ m, i.e., $r_0 = 10^{-6}$ m. The diffusion coefficients are given by $D_{\text{m}} = 0.5 \times 10^{-9}$ m$^2$/s, $D_{\text{tx}} = 1 \times 10^{-10}$ m$^2$/s, and $D_{\text{rx}} = 0.5 \times 10^{-12}$ m$^2$/s. The drift velocity of the fluid medium is considered as $v = 10^{-3}$ m/s. Figs. 3-5 demonstrate the impact of releasing molecules at different time slots $k$ on the first hitting time PDF under various mobility conditions. Moreover, Fig. 6 compares the arrival probabilities $q_0 = \int_0^T f_{T_h}(t;k)\mathrm{d}t$ with $T = 0.3$ ms for various flow velocities $v \in \{10^{-3}, 0.5 \times 10^{-3}\}$ m/s and mobility scenarios. Fig. 3 shows the PDF of the first hitting time for fixed TX and mobile RX. In this case the first hitting time follows a Lévy distribution with the parameters $\bar{d}_k$ and $D_{\text{eff}}$ (cf. Sec. III-B1). Since only the RX is moving, the distance between TX and RX becomes larger as the time slot $k$ increases (cf. (8) and (9)) and, as a result, the probability for a small first hitting time decreases compared to the scenario of fixed TX and RX. This observation is confirmed through the arrival probability $q_0$ shown in Fig. 6, which is almost zero from $k = 2$ and $k = 4$ for $v = 10^{-3}$ m/s and $v = 0.5 \times 10^{-3}$ m/s, respectively.

Fig. 4 shows the PDF of the first hitting time for mobile TX and fixed RX. In this case, first the TX moves towards the fixed RX, and, thus, the distance between them reduces. After a certain time, the TX passes the RX and the distance between them becomes larger. Once the TX has passed the RX, molecular communication with negative flow occurs [25]. This behavior can be observed in Fig. 6, where the arrival probability $q_0$ first increases (TX moves towards RX) and then decreases (TX has passed RX). In particular, an increase in the arrival probability can be observed as long as $\text{sgn}(\bar{d}_k) > 0$,

$$f_{T_h}(t;k) = \frac{\sqrt{kTD_{\text{tx}}D_{\text{m}}}}{\pi(kTD_{\text{tx}}+D_{\text{m}}t)\sqrt{t}} \exp\left(\frac{-D_{\text{m}}\bar{d}_k^2 - kTD_{\text{tx}}(v^\star)^2 t}{4kTD_{\text{tx}}D_{\text{m}}}\right)$$
$$+ \frac{1}{2} f_{T_h}\left(t + \frac{kTD_{\text{tx}}}{D_{\text{m}}}\right) \exp\left(\frac{v^\star(v^\star(kTD_{\text{tx}})^2 + 2v^\star kTD_{\text{tx}}D_{\text{m}}t - 2\bar{d}_k D_{\text{m}}^2 t - 2\bar{d}_k kTD_{\text{tx}}D_{\text{m}})}{4D_{\text{m}}^2(D_{\text{m}}t + kTD_{\text{tx}})}\right)$$
$$\times \sum_{l=1}^{2}\left(\frac{v^\star kTD_{\text{tx}}}{D_{\text{m}}\bar{d}_k} + (-1)^l\right) \exp\left((-1)^l \frac{v^\star \bar{d}_k t}{2(D_{\text{m}}t + kTD_{\text{tx}})}\right) \left[1 + \text{erf}\left(\frac{v^\star kTD_{\text{tx}} + (-1)^l \bar{d}_k D_{\text{m}}}{\sqrt{4kTD_{\text{tx}}D_{\text{m}}}}\sqrt{\frac{t}{D_{\text{m}}t + kTD_{\text{tx}}}}\right)\right]. \quad (15)$$

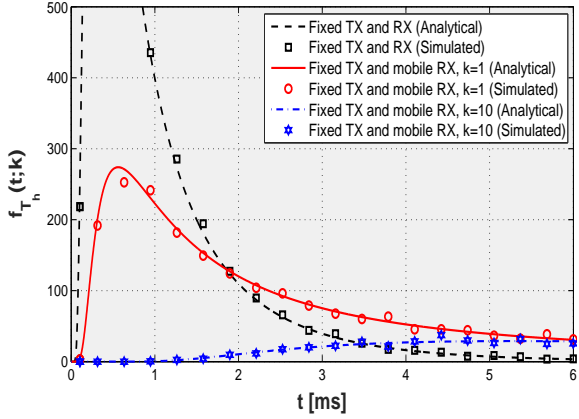

Fig. 3. First hitting time PDF $f_{T_h}(t;k)$ for molecules released at different time slots $k$, with fixed TX and mobile RX.

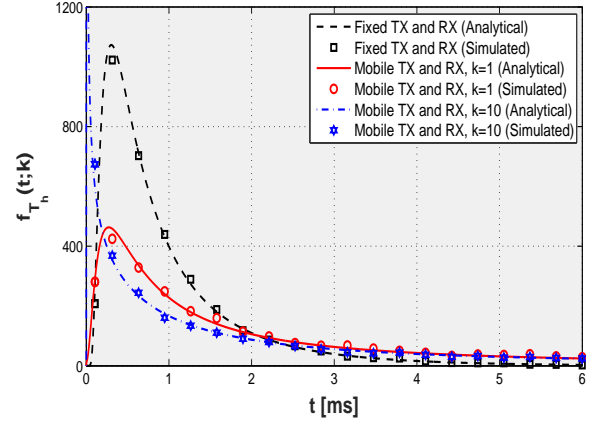

Fig. 5. First hitting time PDF $f_{T_h}(t;k)$ for molecules released at different time slots $k$, with mobile TX and mobile RX.

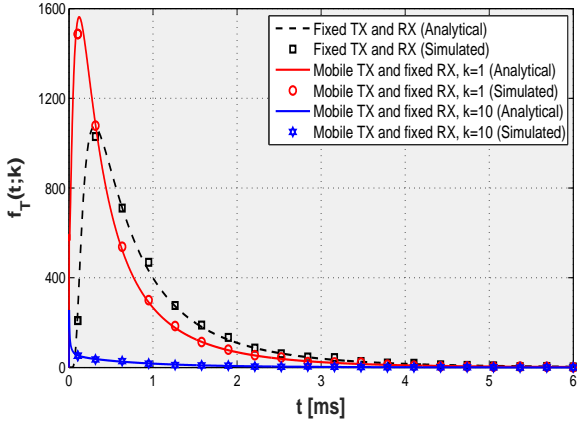

Fig. 4. First hitting time PDF $f_{T_h}(t;k)$ for molecules released at different time slots $k$, with mobile TX and fixed RX.

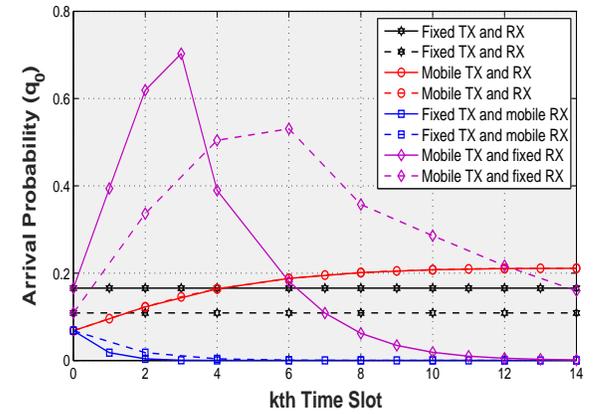

Fig. 6. Arrival probability $q_0 = \int_0^T f_{T_h}(t;k)dt$ for molecules released at time $t = kT$ for $v = 10^{-3}$ m/s (solid line) and $v = 0.5 \times 10^{-3}$ m/s (dashed line).

which is the case for $k \leq 3$ and $k \leq 6$ for $v = 10^{-3}$ m/s and $v = 0.5 \times 10^{-3}$ m/s, respectively.

Fig. 5 shows the PDF of the first hitting time for mobile TX and RX in a diffusion channel with flow $v$. This scenario is identical to the case of mobile TX and RX in a diffusion channel without flow (cf. Sec. III-B3). As the time slot $k$ increases the variation of the distance between TX and RX becomes larger (cf. (8) and (9)), resulting in a non-zero probability that TX and RX are in close proximity. Thus, a non-zero probability for very small hitting times can be observed, as shown in Fig. 5 for $k = 10$. The arrival probability $q_0$ first increases, but ultimately it becomes zero for $k \to \infty$ as shown in [18, Fig. 3].

The analytical expression of the first hitting time PDF is verified through a particle-based simulation. For this purpose we modeled the the movement of TX and RX as well as the the motion of the information molecules as random walk [1] and determined the relative frequency of the first molecule arrivals. As can be seen from Figs. 3-5, the analytical PDF expressions in Section III-B match exactly with the particle-based simulations under aforementioned mobility conditions, thereby validating the derived analytical results.

## IV. PEFORMANCE ANALYSIS

This section comprehensively analyzes the performance of flow-induced diffusive mobile molecular communication system in terms of detection, bit error probability, and capacity considering the impact of ISI, counting errors, and noise from the other sources at the RX. The number of molecules received at the RX during time slot $[(j-1)T, jT]$ can be expressed as

$$R[j] = S[j] + \mathcal{I}[j] + N[j] + C[j]. \quad (17)$$

The quantity $S[j]$ denotes the number of molecules received in the current slot $[(j-1)T, jT]$ and is binomially distributed with parameters $\mathcal{Q}[j]x[j]$ and $q_0$, where $x[j] \in \{0, 1\}$ is the symbol transmitted by the TX in the $j$th time slot. The quantity $N[j]$ denotes the MSI, i.e., noise arising due to molecules received from the other sources, which can be modeled as a Gaussian random variable with mean $\mu_o$ and variance $\sigma_o^2$ assuming a sufficiently large number of interfering sources [29]. Further, the noise $N[j]$ is independent of the number of molecules $S[j]$ and $\mathcal{I}[j]$ received from the intended TX [23]. The term $C[j]$ denotes counting error at the RX and can be modeled as a Gaussian distributed random variable with zero mean and variance $\sigma_c^2[j]$ that depends on the average number of molecules received, i.e., $\sigma_c^2[j] = \mathbb{E}\{R[j]\}$ [23], [30]. The quantity $\mathcal{I}[j]$ models the ISI, arising from transmission in the previous $j-1$ slots, and is determined as

$$\mathcal{I}[j] = I[1] + I[2] + \cdots + I[j-1], \quad (18)$$

where $I[k] \sim Binomial(\mathcal{Q}[j-k]x[j-k], q_k), 1 \leq k \leq j-1$, denotes the number of stray molecules received from the previous $(j-k)$th slot. Assuming the number of molecules released by the TX to be sufficiently large, the binomial distribution for $S[j]$ can be approximated by a Gaussian distribution[2] with mean $\mu[j] = \mathcal{Q}[j]x[j]q_0$ and variance $\sigma^2[j] = \mathcal{Q}[j]x[j]q_0(1-q_0)$, i.e., $S[j] \sim \mathcal{N}(\mathcal{Q}[j]x[j]q_0, \mathcal{Q}[j]x[j]q_0(1-q_0))$ [31]. Similarly, the binomial distribution for $I[k], 1 \leq k \leq j-1$ can be approximated as $I[k] \sim \mathcal{N}(\mu_I[k] = \mathcal{Q}[j-k]x[j-k]q_k, \sigma_I^2[k] = \mathcal{Q}[j-k]x[j-k]q_k(1-q_k))$. Further, it can be noted that $S[j]$ and $I[k], k = 1, 2, \cdots, j-1$ are mutually independent since the molecules transmitted in different time slots do not interfere with each other [23], [24].

### A. Optimal Decision Rule at the RX

The symbol detection problem at the RX can be formulated as the binary hypothesis testing problem

$$\begin{aligned}\mathcal{H}_0 : R[j] =& I[1] + I[2] + \cdots + I[j-1] + N[j] + C[j]\\ \mathcal{H}_1 : R[j] =& S[j] + I[1] + I[2] + \cdots + I[j-1]\\ &+ N[j] + C[j], \end{aligned} \quad (19)$$

where the null and alternative hypotheses $\mathcal{H}_0, \mathcal{H}_1$ correspond to the transmission of binary symbols 0, 1 respectively during the $j$th time slot. The number of molecules $R[j]$ received at the RX corresponding to the different hypotheses are distributed as

$$\begin{aligned}\mathcal{H}_0 &: R[j] \sim \mathcal{N}(\mu_0[j], \sigma_0^2[j])\\ \mathcal{H}_1 &: R[j] \sim \mathcal{N}(\mu_1[j], \sigma_1^2[j]), \end{aligned} \quad (20)$$

where the mean $\mu_0[j]$ and variance $\sigma_0^2[j]$ under hypothesis $\mathcal{H}_0$, as derived in Appendix B, are given as

$$\begin{aligned}\mu_0[j] =& \mu_I[1] + \mu_I[2] + \cdots + \mu_I[j-1] + \mu_o\\ =& \beta \sum_{k=1}^{j-1} \mathcal{Q}[j-k]q_k + \mu_o, \end{aligned} \quad (21)$$

$$\begin{aligned}\sigma_0^2[j] =& \sigma_I^2[1] + \sigma_I^2[2] + \cdots + \sigma_I^2[j-1] + \sigma_o^2 + \sigma_c^2[j]\\ =& \sum_{k=1}^{j-1}[\beta\mathcal{Q}[j-k]q_k(1-q_k) + \beta(1-\beta)\\ &\times (\mathcal{Q}[j-k]q_k)^2] + \sigma_o^2 + \mu_0[j], \end{aligned} \quad (22)$$

and the mean $\mu_1[j]$ and variance $\sigma_1^2[j]$ under hypothesis $\mathcal{H}_1$, as derived in Appendix C, are given as

$$\begin{aligned}\mu_1[j] =& \mu[j] + \mu_I[1] + \mu_I[2] + \cdots + \mu_I[j-1] + \mu_o\\ =& \mathcal{Q}[j]q_0 + \beta \sum_{k=1}^{j-1} \mathcal{Q}[j-k]q_k + \mu_o, \end{aligned} \quad (23)$$

$$\begin{aligned}\sigma_1^2[j] =& \sigma^2[j] + \sigma_I^2[1] + \sigma_I^2[2] + \cdots + \sigma_I^2[j-1] + \sigma_o^2 + \sigma_c^2[j]\\ =& \mathcal{Q}[j]q_0(1-q_0) + \sum_{k=1}^{j-1}[\beta\mathcal{Q}[j-k]q_k(1-q_k)\\ &+ \beta(1-\beta)(\mathcal{Q}[j-k]q_k)^2] + \sigma_o^2 + \mu_1[j]. \end{aligned} \quad (24)$$

The optimal decision rule at the RX for symbol detection is presented next.

**Theorem 2.** *The optimal decision rule at the RX corresponding to the $j$th time slot is obtained as*

$$T(R[j]) = R[j] \underset{\mathcal{H}_0}{\overset{\mathcal{H}_1}{\gtrless}} \gamma'[j], \quad (25)$$

*where the optimal decision threshold $\gamma'[j]$ is given as*

$$\gamma'[j] = \sqrt{\gamma[j]} - \alpha[j]. \quad (26)$$

*The quantities $\gamma[j]$ and $\alpha[j]$ are defined as*

$$\alpha[j] = \frac{\mu_1[j]\sigma_0^2[j] - \mu_0[j]\sigma_1^2[j]}{\sigma_1^2[j] - \sigma_0^2[j]}, \quad (27)$$

$$\begin{aligned}\gamma[j] =& \frac{2\sigma_1^2[j]\sigma_0^2[j]}{\sigma_1^2[j] - \sigma_0^2[j]} \ln\left[\frac{(1-\beta)}{\beta}\sqrt{\frac{\sigma_1^2[j]}{\sigma_0^2[j]}}\right] + (\alpha[j])^2\\ &+ \frac{\mu_1^2[j]\sigma_0^2[j] - \mu_0^2[j]\sigma_1^2[j]}{\sigma_1^2[j] - \sigma_0^2[j]}. \end{aligned} \quad (28)$$

*Proof.* The optimal log likelihood ratio test (LLRT) at the RX is given as

$$\Lambda(R[j]) = \ln\left[\frac{p(R[j]|\mathcal{H}_1)}{p(R[j]|\mathcal{H}_0)}\right] \underset{\mathcal{H}_0}{\overset{\mathcal{H}_1}{\gtrless}} \ln\left[\frac{1-\beta}{\beta}\right]. \quad (29)$$

---

[2]This approximation is reasonable when $\mathcal{Q}[j]q_0 > 5$ and $\mathcal{Q}[j](1-q_0) > 5$ [23].

Substituting the Gaussian PDFs $p(R[j]|\mathcal{H}_1)$ and $p(R[j]|\mathcal{H}_0)$ from (20), the test statistic $\Lambda(R[j])$ can be obtained as

$$\Lambda(R[j]) = \ln\left[\sqrt{\frac{\sigma_0^2[j]}{\sigma_1^2[j]}}\right] + \frac{1}{2\sigma_0^2[j]\sigma_1^2[j]}$$
$$\times \underbrace{(R[j]-\mu_0[j])^2\sigma_1^2[j] - (R[j]-\mu_1[j])^2\sigma_0^2[j]}_{\triangleq f(R[j])}. \quad (30)$$

The expression for $f(R[j])$ given above can be further simplified as

$$f(R[j]) = R^2[j](\sigma_1^2[j]-\sigma_0^2[j]) + 2R[j](\mu_1[j]\sigma_0^2[j]$$
$$- \mu_0[j]\sigma_1^2[j]) + (\mu_0^2[j]\sigma_1^2[j]-\mu_1^2[j]\sigma_0^2[j])$$
$$= (\sigma_1^2[j]-\sigma_0^2[j])(R[j]+\alpha[j])^2$$
$$- \frac{(\mu_1[j]\sigma_0^2[j]-\mu_0[j]\sigma_1^2[j])^2}{\sigma_1^2[j]-\sigma_0^2[j]}$$
$$+ (\mu_0^2[j]\sigma_1^2[j]-\mu_1^2[j]\sigma_0^2[j]), \quad (31)$$

where $\alpha[j]$ is defined in (27). Substituting the above equation for $f(R[j])$ in (30) and subsequently merging the terms independent of the received molecules $R[j]$ with the detection threshold, the test can be equivalently expressed as

$$(R[j]+\alpha[j])^2 \underset{\mathcal{H}_0}{\overset{\mathcal{H}_1}{\gtrless}} \gamma[j], \quad (32)$$

where $\gamma[j]$ is defined in (28). Further, taking the square root of both sides where $\gamma[j] \geq 0$, Equation (32) can be simplified to yield the optimal test in (25). $\square$

### B. Probabilities of Detection and False Alarm

The detection performance for the optimal test derived in (25) at the RX considering also the mobility of the TX and RX, is obtained next.

**Theorem 3.** *The average probabilities of detection $(P_D)$ and false alarm $(P_{FA})$ at the RX in the diffusion based mobile molecular communication nano-network, corresponding to the transmission by the TX over $i$ slots, are given as*

$$P_D = \frac{1}{i}\sum_{j=1}^{i} P_D[j]$$
$$= \frac{1}{i}\sum_{j=1}^{i} Q\left(\frac{\gamma'[j]-\mu_1[j]}{\sigma_1[j]}\right), \quad (33)$$
$$P_{FA} = \frac{1}{i}\sum_{j=1}^{i} P_{FA}[j]$$
$$= \frac{1}{i}\sum_{j=1}^{i} Q\left(\frac{\gamma'[j]-\mu_0[j]}{\sigma_0[j]}\right), \quad (34)$$

*where $\gamma'[j]$ is defined in (26) and $Q(\cdot)$ denotes the tail probability of the standard normal random variable [32].*

*Proof.* The probabilities of detection $(P_D[j])$ and false alarm $(P_{FA}[j])$ at the RX in the $j$th time slot for the decision rule in (25) are obtained as

$$P_D[j] = \Pr(T(R[j]) > \gamma'[j]|\mathcal{H}_1)$$
$$= \Pr(R[j] > \gamma'[j]|\mathcal{H}_1), \quad (35)$$
$$P_{FA}[j] = \Pr(T(R[j]) > \gamma'[j]|\mathcal{H}_0)$$
$$= \Pr(R[j] > \gamma'[j]|\mathcal{H}_0), \quad (36)$$

where the number of received molecules $R[j]$ is Gaussian distributed (20) under hypotheses $\mathcal{H}_0$ and $\mathcal{H}_1$ respectively. Subtracting their respective means followed by division by the standard deviations, i.e., $\frac{R[j]-\mu_1[j]}{\sigma_1[j]}$ and $\frac{R[j]-\mu_0[j]}{\sigma_0[j]}$ yields standard normal random variables for hypotheses $\mathcal{H}_1$ and $\mathcal{H}_0$ respectively. Subsequently, the expressions for $P_D[j]$ and $P_{FA}[j]$, given in (33) and (34) respectively, can be obtained employing the definition of the $Q(\cdot)$ function. $\square$

### C. Bit Error Probability

The end-to-end probability of error for communication between TX and RX follows as described in the result below.

**Lemma 1.** *The average probability of error $(P_e)$ for slots $1$ to $i$ at the RX in the diffusion based molecular nano-network with mobile TX and RX is*

$$P_e = \frac{1}{i}\sum_{j=1}^{i}\left[\beta\left(1-Q\left(\frac{\gamma'[j]-\mu_1[j]}{\sigma_1[j]}\right)\right)\right.$$
$$\left.+(1-\beta)Q\left(\frac{\gamma'[j]-\mu_0[j]}{\sigma_0[j]}\right)\right]. \quad (37)$$

*Proof.* The probability of error $P_e[j]$ in the $j$th time slot is defined as [32]

$$P_e[j] = \Pr(\text{decide } \mathcal{H}_0, \mathcal{H}_1 \text{ true}) + \Pr(\text{decide } \mathcal{H}_1, \mathcal{H}_0 \text{ true})$$
$$= (1-P_D[j])P(\mathcal{H}_1) + P_{FA}[j]P(\mathcal{H}_0), \quad (38)$$

where the prior probabilities of the hypotheses $P(\mathcal{H}_1)$ and $P(\mathcal{H}_0)$ are $\beta$ and $1-\beta$ respectively. The quantities $P_D[j]$ and $P_{FA}[j]$ denote the probabilities of detection and false alarm at the RX during the $j$th time slot as obtained in (33) and (34) respectively. The average probability of error for slots $1$ to $i$ follows as stated in (37). $\square$

### D. Capacity

Let $X[j]$ and $Y[j]$ be two discrete random variables that represent the transmitted and received symbols, respectively, in the $j$th slot. The mutual information $I(X[j], Y[j])$ between $X[j]$ and $Y[j]$ with marginal probabilities $\Pr(x[j]=0) = 1-\beta, \Pr(x[j]=1) = \beta$ is given by

$$I(X[j],Y[j]) = \sum_{x[j]\in\{0,1\}}\sum_{y[j]\in\{0,1\}} \Pr(y[j]|x[j])\Pr(x[j])$$
$$\times \log_2 \frac{\Pr(y[j]|x[j])}{\sum_{x[j]\in\{0,1\}}\Pr(y[j]|x[j])\Pr(x[j])}, \quad (39)$$

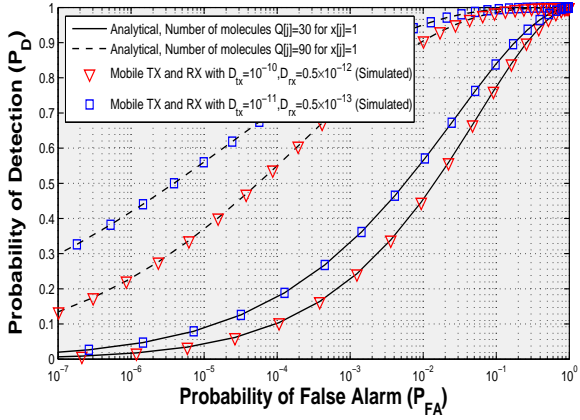

Fig. 7. Probability of detection versus probability of false alarm considering mobile TX and RX scenario with $D_{\text{tx}} \in \{10^{-10}, 10^{-11}\}$ m$^2$/s, $D_{\text{rx}} \in \{0.5 \times 10^{-12}, 0.5 \times 10^{-13}\}$ m$^2$/s, $v_{\text{tx}} = v_{\text{rx}} = v = 10^{-3}$ m/s, and $T = 2$ ms.

where the conditional probabilities $\Pr(y[j] \in \{0,1\}|x[j] \in \{0,1\})$ can be written in terms of $P_D[j]$ and $P_{FA}[j]$ as

$$\Pr(y[j] = 0|x[j] = 0) = 1 - P_{FA}[j]$$
$$= 1 - Q\left(\frac{\gamma'[j] - \mu_0[j]}{\sigma_0[j]}\right),$$
$$\Pr(y[j] = 1|x[j] = 0) = P_{FA}[j]$$
$$= Q\left(\frac{\gamma'[j] - \mu_0[j]}{\sigma_0[j]}\right),$$
$$\Pr(y[j] = 0|x[j] = 1) = 1 - P_D[j]$$
$$= 1 - Q\left(\frac{\gamma'[j] - \mu_1[j]}{\sigma_1[j]}\right),$$
$$\Pr(y[j] = 1|x[j] = 1) = P_D[j]$$
$$= Q\left(\frac{\gamma'[j] - \mu_1[j]}{\sigma_1[j]}\right).$$

The mutual information between the TX and RX can be maximized as

$$C[i] = \max_{\beta} \frac{1}{i} \sum_{j=1}^{i} I(X[j], Y[j]) \text{ bits/slot,} \quad (40)$$

which equals the channel capacity as the number of slots $i \to \infty$ [22].

## V. SIMULATION RESULTS

This section presents simulation results to demonstrate the performance of the diffusive mobile molecular communication system under various mobile conditions and also to validate the derived analytical results. For simulation purposes, the various system parameters are set as, diffusion coefficient $D_m = 5 \times 10^{-10}$ m$^2$/s, slot duration $T \in \{1, 2, 10\}$ ms, distance $r_0 = 1$ μm, prior probability $\beta = 0.5$, and $i = 10$ slots. The MSI at the RX is modeled as a Gaussian distributed random variable with mean $\mu_o = 10$ and variance $\sigma_o^2 = 10$ unless otherwise stated.

Figs. 7-9 demonstrate the detection performance of optimal LRT based detector in (25) under various mobile scenarios with and without flow, where each point on the curve corresponds to a value of $(P_{FA}, P_D)$ for a given threshold[3] $\gamma'[j] = \gamma', \forall j$ [32]. For these simulations, the range of $\gamma'$ is considered from 1 to 80. Fig. 7 demonstrates the detection performance at the RX considering the mobile nature of the TX and RX under drift channel with $v = 10^{-3}$ m/s. It can be clearly seen that an increase in the number of molecules emitted by the TX results in a higher probability of detection at the RX for a fixed value of the probability of false alarm. For example with $D_{\text{tx}} = 10^{-10}$ m$^2$/s and $D_{\text{rx}} = 0.5 \times 10^{-12}$ m$^2$/s, the probability of detection $P_D$ can be increased from 0.1 to 0.52 as the number of molecules $\mathcal{Q}[j]$ released for symbol $x[j] = 1$ increases from 30 to 90 for a fixed probability of false alarm $P_{FA} = 10^{-4}$. On the other hand, the detection performance significantly deteriorates as the diffusion coefficients $D_{\text{tx}}$ and $D_{\text{rx}}$ increase due to higher mobility.

In Figs. 8(a)-(c), one can observe that the detection performance at the RX, considering $\mathcal{Q}[j] = 30$ molecules released by the TX for information symbol $x[j] = 1$, significantly improves as slot duration ($T$) increases from 1 ms to 10 ms. Further, the detection performance at the RX considering the mobility of the TX and RX with drift $v = 10^{-3}$ m/s and without drift $v = 0$ are identical. This is because of the fact that the molecule arrival probabilities $q_{j-k}$ (cf. (1) and (16)) are equivalent under both the scenarios. Furthermore, the system with fixed TX and RX under drift, i.e., $v = 10^{-3}$ m/s outperforms the other scenarios. However, without drift, i.e., $v = 0$ as shown in Fig. 8(a), fixed TX and RX scenario performs poor for $T = 1$ ms in comparison to the mobile TX and RX case. This is owing to the fact that the probability of a molecule reaching the RX within current slot, i.e., $q_0$ is lower and the probabilities of a molecules received from the previous slots, i.e., $q_{j-k}, j \neq k$, are higher under the fixed TX and RX scenario in comparison to the one obtained for mobile TX and RX scenario. On contrary, as $T$ increases from 1 ms to 2 ms and 10 ms shown in Figs. 8(b)-(c), the system where a fixed TX communicates with a fixed RX under no drift outperforms the mobile scenario.

Figs. 9(a) and 9(b) demonstrate the detection performance at the RX under fixed TX and mobile RX, and mobile TX and fixed RX scenarios respectively, where the following interesting observations are obtained. First, the detection performance considering fixed TX and mobile RX scenario significantly degrades under drift channel and the performance further deteriorates as the drift velocity ($v$) increases from $5 \times 10^{-5}$ m/s to $2 \times 10^{-4}$ m/s. This is due to the fact that the mobility of RX under drift channel increases the distance between TX and RX that in turn decreases the probability of a molecule arriving in same slot and also increase the ISI with higher arrival probabilities from previous slots. This can also be clearly seen in Fig. 3 with $k = 10$ slots. Under the scenario with mobile TX and fixed RX as shown in Fig. 9(b), the detection performance at the RX with drift velocity $v = 10^{-4}$ m/s improves in comparison to the scenario with no drift, i.e., $v = 0$. This arises due to the fact that the mobile

---
[3]It is important to note that this is not an optimal threshold derived in (26).

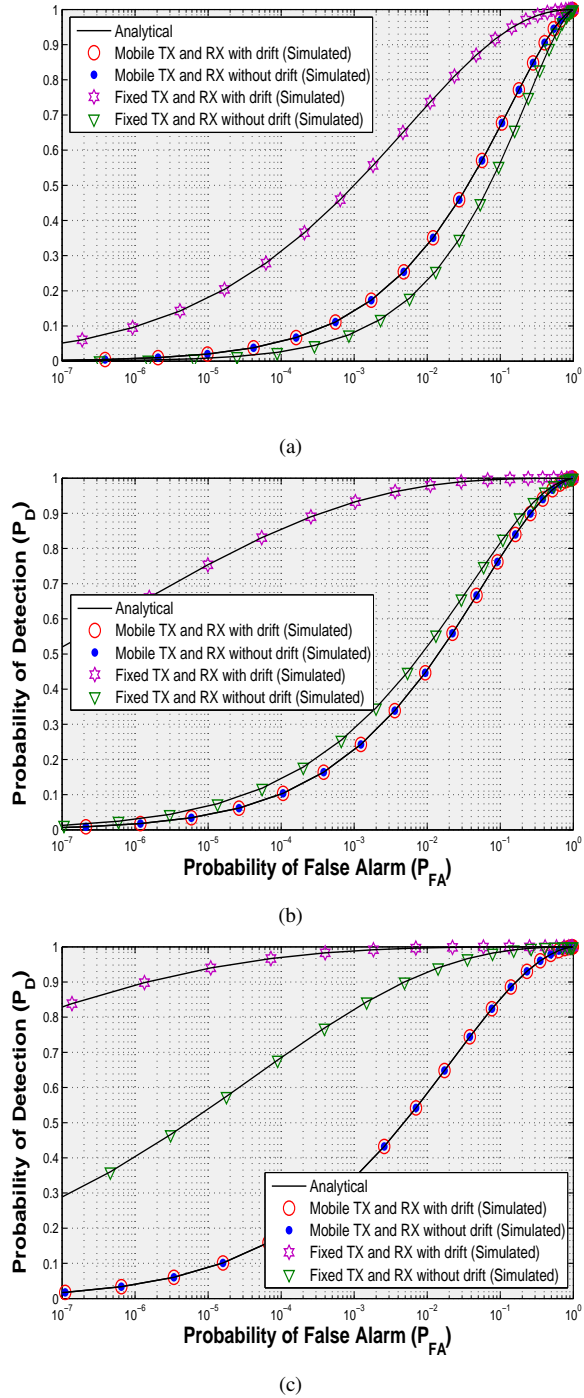

Fig. 8. Probability of detection versus probability of false alarm considering mobile TX and RX scenario with $D_{\text{tx}} = 10^{-10}$ m$^2$/s, $D_{\text{rx}} = 0.5 \times 10^{-12}$ m$^2$/s, and fixed TX and RX scenario with $D_{\text{tx}} = D_{\text{rx}} = v_{\text{tx}} = v_{\text{rx}} = 0$ over slot duration (a) $T = 1$ ms (b) $T = 2$ ms, and (c) $T = 10$ ms.

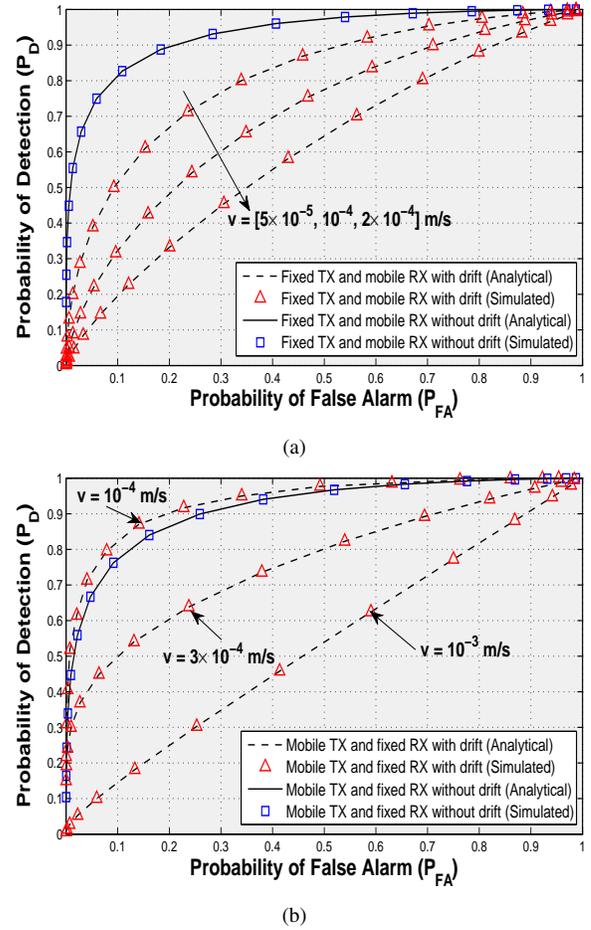

Fig. 9. Probability of detection versus probability of false alarm considering $\mathcal{Q}[j] = 30$ molecules released by the TX for information symbol $x[j] = 1$, slot duration $T = 2$ ms, and (a) fixed TX and mobile RX scenario with $D_{\text{tx}} = 0$, $D_{\text{rx}} = 0.5 \times 10^{-12}$ m$^2$/s, $v_{\text{tx}} = 0, v_{\text{rx}} = v$ m/s and (b) mobile TX and fixed RX with $D_{\text{tx}} = 10^{-10}$ m$^2$/s, $D_{\text{rx}} = 0$, $v_{\text{tx}} = v$ m/s, $v_{\text{rx}} = 0$.

TX with $v = 10^{-4}$ m/s under drift channel approaches the RX that in turn increases the probability $q_0$ as slot number $j$ progressively increases. However, as drift velocity increases from $10^{-4}$ m/s to $3 \times 10^{-4}$ m/s and $10^{-3}$ m/s, the TX crosses the RX and the distance between them further increases as slot number $j$ increases. Therefore, in comparison to a scenario with mobile TX and fixed RX with drift $v \in \{3 \times 10^{-4}, 10^{-3}\}$ m/s, a significant performance improvement can be seen for a scenario when mobile TX communicates with fixed RX over diffusive channel without drift.

Fig. 10 presents the error rate versus MSI noise variance ($\sigma_o^2$) performance under various mobile scenarios, where the optimal decision rule (25) is employed at the RX. First, it can be observed from Fig. 10 that the analytical $P_e$ values obtained using (37) coincide with those obtained from simulations, thus validating the derived analytical results. One can also observe that an increase in the MSI noise variance ($\sigma_o^2$) results in a higher probability of error at the RX. The error rate further increases as the mobility increases as shown in Fig. 10(a). Moreover, it can also be seen therein that the system with $T = 10$ ms experiences lower probability of error for the fixed TX and RX scenario with and without drift in comparison to the mobile scenario. Further, similar to detection performance, the system considering mobile TX and fixed RX with flow $v = 10^{-4}$ m/s experiences lower probability of error in comparison to the scenario when both mobile TX and fixed RX communicate with no flow $v = 0$. On the other hand, for fixed TX and mobile RX as shown in Fig. 10(b), the system performance significantly degrades under drift channel and the performance further deteriorates as the drift velocity

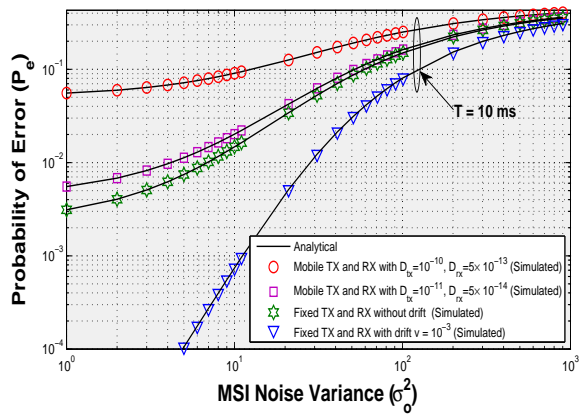

(a)

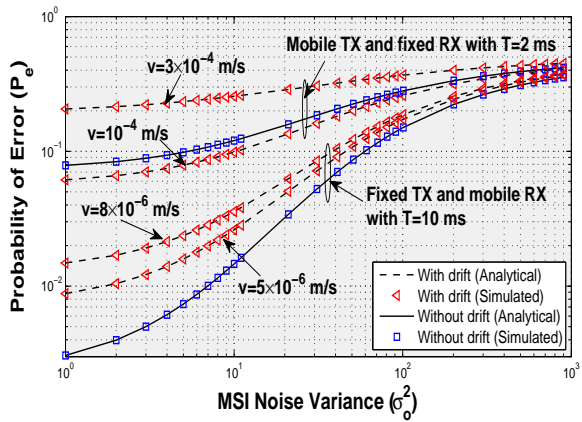

(b)

Fig. 10. Probability of error versus MSI noise variance ($\sigma_o^2$) with $\mu_o = 0$ and $\mathcal{Q}[j] = 30$ molecules released by the TX for information symbol $x[j] = 1$.

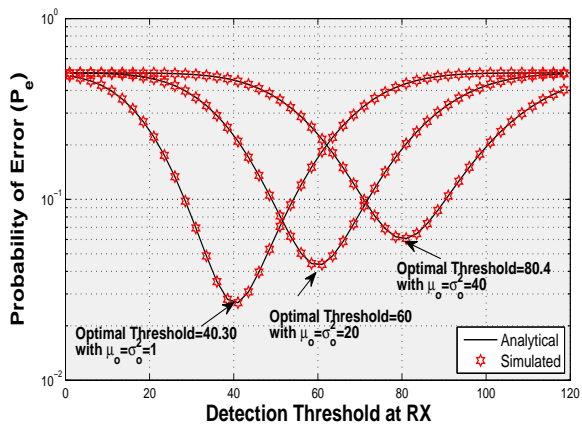

Fig. 11. Error rate of diffusion based molecular systems at $j = 10$ time slot versus detection threshold for mobile TX and RX scenario with $\mathcal{Q}[j] = 120$ molecules released by the TX for information symbol $x[j] = 1$.

$v$ increases.

Fig. 11 shows the error rate performance at $(j = 10)$th time slot versus detection threshold at the RX considering mobile TX and RX with diffusion coefficients $D_{\text{tx}} = 10^{-10}$ m$^2$/s, $D_{\text{rx}} = 0.5 \times 10^{-12}$ m$^2$/s and slot duration $T = 10$ ms. It can be seen that the optimal threshold obtained using (26) is able to achieve the minimum probability of error at the RX. The optimal value of the threshold (cf. (26)) as shown in Fig. 11 increases from 40.30 to 80.4 as the MSI increases from $\mu_o = \sigma_o^2 = 1$ to $\mu_o = \sigma_o^2 = 40$.

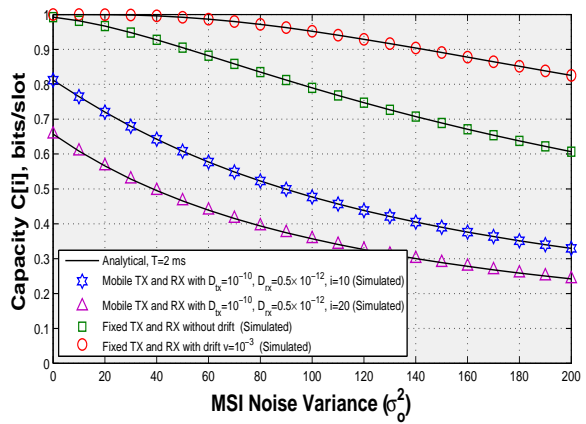

(a)

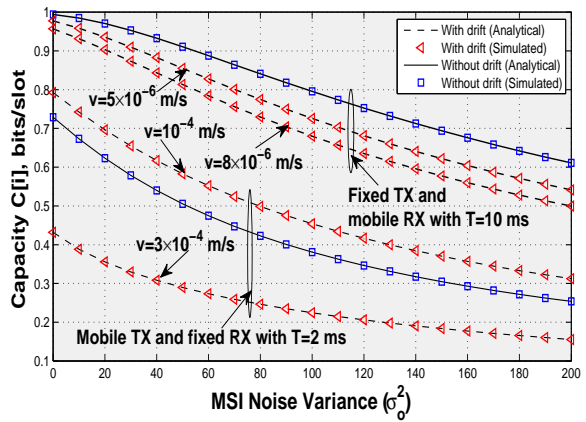

(b)

Fig. 12. Capacity versus MSI noise variance ($\sigma_o^2$) under various mobile conditions with $\mu_o = 10$ and $\mathcal{Q}[j] = 60$ molecules released by the TX for information symbol $x[j] = 1$.

Fig. 12 shows the capacity performance where the maximum mutual information is achieved for equiprobable information symbols, i.e., $\beta = 0.5$. Similar to the error rate performance, the maximum mutual information $C[i]$ obtained for slots 1 to $i$ also decreases as the MSI noise variance ($\sigma_o^2$) increases. Interestingly, it is also worth noting that the maximum mutual information $C[i]$ in bits per channel use progressively decreases as the number of slots $(i)$ increases due to the mobile nature of TX and RX. This is owing to the fact that the probability of a molecule reaching the RN within the current slot, i.e., $q_0$ progressively decreases while the ISI from previous slots increases as the value of $i$ increases. Further, the value of $C[i]$ considering mobile TX and fixed RX with flow $v = 10^{-4}$ m/s as shown in Fig. 12(b) is higher compared to the scenario when both mobile TX and fixed RX communicate over a diffusive channel without drift.

## VI. CONCLUSION

Closed-form expressions for the PDF of the first hitting time in flow-induced diffusive molecular communication under various mobile scenarios were derived. Based on these results the performance for molecular communication in flow-induced diffusion channels in the presence of ISI, MSI, and counting errors was investigated. Closed-form analytical expressions were derived for the optimal decision test statistics and optimal decision threshold, together with the resulting probabilities of detection, false alarm as well as the probability of error. In addition, an analytical expression for the capacity was derived. Simulation results were conducted to verify the theoretical results and to gain insights into the performance under various mobility conditions. The results reveal that molecular communication of mobile TX and RX in a drift channel performs identical to mobile TX and RX in a non-drift environment. For fixed TX and moving RX, the error performance degrades for an increasing drift velocity, since the distance between TX and RX increases faster. In case of mobile TX and fixed RX, the error performance improves as long as TX moves towards RX, but significantly deteriorates when TX has passed RX because molecular communication with negative flow occurs in this case. Finally, future studies can focus on the first hitting time distribution in 3D environment. However, this task is not straightforward since, to the best of our knowledge, no closed-form expression for the first hitting time is known even for fixed TX and RX and a fully absorbing RX.

## APPENDIX A
## PROOF OF (15)

Substituting (9) and (14) into (13) results in

$$f_{T_h}(t;k) = \frac{1}{\sqrt{4\pi D_m t^3}} \frac{1}{\sqrt{\sigma_k^2}} \sqrt{\frac{\lambda}{\sqrt{\sigma_k^2}}} \exp\left(-\frac{\lambda^2}{2}\right)$$
$$\times \int_{r_k=0}^{\infty} r_k^{3/2} \exp\left(-\frac{(v^\star t - r_k)^2}{4 D_m t}\right) \exp\left(-\frac{r_k^2}{2\sigma_k^2}\right) I_{-\frac{1}{2}}\left(\frac{\lambda r_k}{\sqrt{\sigma_k^2}}\right) dr_k$$
(41)

The integral in (41) can be written as

$$Z[k] = \int_{r_k=0}^{\infty} r_k^{3/2} \exp\left(-\frac{(v^\star t - r_k)^2}{4 D_m t}\right) \exp\left(-\frac{r_k^2}{2\sigma_k^2}\right) I_{-\frac{1}{2}}\left(\frac{\lambda r_k}{\sqrt{\sigma_k^2}}\right) dr_k$$
$$= \int_{r_k=0}^{\infty} r_k^{3/2} \exp(-a(b-r_k)^2) \exp(-c r_k^2) I_{-\frac{1}{2}}(d r_k) dr_k$$
$$= \exp(-(ab)^2)$$
$$\times \int_{r_k=0}^{\infty} r_k^{3/2} \exp(-(a+c)r_k^2) \exp(2ab r_k) I_{-\frac{1}{2}}(d r_k) dr_k,$$
(42)

with $a = \frac{1}{4 D_m t}$, $b = v^\star t$, $c = \frac{1}{2\sigma_k^2}$ and $d = \frac{\lambda}{\sqrt{\sigma_k^2}}$. Applying the relation $I_{-\frac{1}{2}}(y) = \sqrt{\frac{1}{2\pi y}}[\exp(y) + \exp(-y)]$ gives

$$Z[k] = \frac{\exp(-(ab)^2)}{\sqrt{2\pi d}} \sum_{l=1}^{2} \int_{r_k=0}^{\infty} r_k \exp(-(a+c)r_k^2)$$
$$\times \exp((2ab + (-1)^l d) r_k) dr_k.$$
(43)

Further, using the identity [26, Eq. (3.462.5)]

$$\int_0^\infty x \exp(-ux^2 - 2wx) dx$$
$$= \frac{1}{2u} - \frac{w}{2u}\sqrt{\frac{\pi}{u}} \exp\left(\frac{w^2}{u}\right) \left(1 - \mathrm{erf}\left(\frac{w}{\sqrt{u}}\right)\right),$$
(44)

leads to

$$Z[k]$$
$$= \mathcal{C} \sum_{l=1}^{2}\left[\frac{1}{2u} - \frac{w_l}{2u}\sqrt{\frac{\pi}{u}} \exp\left(\frac{(w_l)^2}{u}\right)\left(1 - \mathrm{erf}\left(\frac{w_l}{\sqrt{u}}\right)\right)\right],$$
$$= \mathcal{C}\left[\frac{1}{u} - \frac{1}{2u}\sqrt{\frac{\pi}{u}} \sum_{l=1}^{2} w_l \exp\left(\frac{(w_l)^2}{u}\right)\left(1 - \mathrm{erf}\left(\frac{w_l}{\sqrt{u}}\right)\right)\right],$$
(45)

with $\mathcal{C} = \exp(-(ab)^2)/\sqrt{2\pi d}$, $w_l = (2ab + (-1)^l d)$ and $u = a + c$. Re-substituting $w_l$ and $u$ into (45) and $Z[k]$ into (41) results, after some straightforward mathematical manipulations, in (15).

## APPENDIX B
## MEAN $\mu_0[j]$ AND VARIANCE $\sigma_0^2[j]$ UNDER HYPOTHESIS $\mathcal{H}_0$

Using (19), the mean $\mu_0[j]$ under $\mathcal{H}_0$ can be calculated as

$$\mu_0[j] = \mathbb{E}\left\{\sum_{k=1}^{j-1} I[k] + N[j] + C[j]\right\}$$
$$= \sum_{k=1}^{j-1} \mathbb{E}\{I[k]\} + \mu_o,$$
(46)

where $\mathbb{E}\{I[k]\}$ is given as

$$\mathbb{E}\{I[k]\} = \Pr(x[j-k]=1)\mathbb{E}\{I[k]|x[j-k]=1\}$$
$$+ \Pr(x[j-k]=0)\mathbb{E}\{I[k]|x[j-k]=0\}$$
$$= \beta \mathbb{E}\{I[k]|x[j-k]=1\} + (1-\beta)\mathbb{E}\{I[k]|x[j-k]=0\}$$
$$= \beta \mathcal{Q}[j-k] q_k.$$
(47)

The variance $\sigma_0^2[j]$ under $\mathcal{H}_0$ can be derived as

$$\sigma_0^2[j] = \sum_{k=1}^{j-1} \sigma_I^2[k] + \sigma_o^2 + \sigma_c^2[j]$$
$$= \sum_{k=1}^{j-1} \sigma_I^2[k] + \sigma_o^2 + \mu_0[j],$$
(48)

where $\sigma_c^2[j] = \mu_0[j]$ and the variance $\sigma_I^2[k]$ of the ISI term can be obtained as

$$\sigma_I^2[k] = \mathbb{E}\{(I[k])^2\} - \mathbb{E}^2\{I[k]\}$$
$$= \mathbb{E}\{(I[k])^2\} - (\beta \mathcal{Q}[j-k] q_k)^2,$$
(49)

where $\mathbb{E}\{(I[k])^2\}$ is given as

$$\mathbb{E}\{(I[k])^2\} = \Pr(x[j-k]=1)\mathbb{E}\{(I[k])^2|x[j-k]=1\}$$
$$+ \Pr(x[j-k]=0)\mathbb{E}\{(I[k])^2|x[j-k]=0\}$$
$$= \beta \mathbb{E}\{(I[k])^2|x[j-k]=1\}$$
$$= \beta\left[\mathcal{Q}[j-k] q_k(1-q_k) + (\mathcal{Q}[j-k] q_k)^2\right].$$
(50)

Substituting the above expression in (49), the final expression for the variance $\sigma_I^2[k]$ of ISI term is given as

$$\sigma_I^2[k] = \beta \mathcal{Q}[j-k] q_k (1-q_k) + \beta(1-\beta)(\mathcal{Q}[j-k]q_k)^2. \quad (51)$$

## Appendix C
### Mean $\mu_1[j]$ and Variance $\sigma_1^2[j]$ under Hypothesis $\mathcal{H}_1$

Similar to $\mu_0[j]$, the mean $\mu_1[j]$ under $\mathcal{H}_1$ can be calculated using (19) as

$$\begin{aligned}
\mu_1[j] &= \mathbb{E}\left\{S[j] + \sum_{k=1}^{j-1} I[k] + N[j] + C[j]\right\} \\
&= \mathbb{E}\{S[j]\} + \sum_{k=1}^{j-1} \mathbb{E}\{I[k]\} + \mu_o \\
&= \mathcal{Q}[j] q_0 + \beta \sum_{k=1}^{j-1} \mathcal{Q}[j-k] q_k + \mu_0. \quad (52)
\end{aligned}$$

The variance $\sigma_1^2[j]$ under $\mathcal{H}_1$ can be derived as

$$\begin{aligned}
\sigma_1^2[j] &= \sigma^2[j] + \sum_{k=1}^{j-1} \sigma_I^2[k] + \sigma_o^2 + \sigma_c^2[j] \\
&= \sigma^2[j] + \sum_{k=1}^{j-1} \sigma_I^2[k] + \sigma_o^2 + \mu_1[j], \quad (53)
\end{aligned}$$

where $\sigma^2[j] = \mathcal{Q}[j] q_0 (1-q_0)$, $\sigma_c^2[j] = \mu_1[j]$, and $\sigma_I^2[k]$ is given in (51).